\documentclass{iau} 
\usepackage{graphicx}

\title[Spectroscopy of candidate GW EM counterparts] 
{Spectroscopy of candidate electromagnetic counterparts to gravitational wave sources}

\author[I. A. Steele, C. M. Copperwheat, A. S. Piascik]   
{Iain A. Steele, Chris M. Copperwheat \and Andrzej S. Piascik}

\affiliation{Astrophysics Research Institute, Liverpool John Moores University, L3 5RF, UK}

\pubyear{2017}
\volume{324}  
\jname{New Frontiers in Black Hole Astrophysics}
\editors{A.C. Editor, B.D. Editor \& C.E. Editor, eds.}
\begin{document}

\maketitle

\begin{abstract}
A programme of worldwide, multi-wavelength electromagnetic follow-up of sources
detected by gravitational wave detectors is in place.  Following the discovery of
GW150914 and GW151226, wide field imaging of their sky localisations 
identified a number of candidate optical counterparts which were then
spectrally classified.  The majority of 
candidates were found to be supernovae at redshift ranges similar to the GW events and were
thereby ruled out as a genuine counterpart.  Other candidates ruled out include AGN and Solar System
objects.  Given the GW sources were black hole binary mergers, the lack of an
identified electromagnetic counterpart is not surprising.  However the observations show that is
it is possible to organise and execute a campaign that can eliminate the majority
of potential counterparts.  Finally 
we note the existence of a ``classification gap'' with a significant fraction of 
candidates going unclassified. 
\end{abstract}

\firstsection 
\section{Introduction}

As part of the effort to discover and characterise astronomical gravitational wave (GW)
sources, a worldwide programme of electromagnetic (EM) follow-up has been established 
(\cite[Abbot et al. 2016a]{abbott-a}).  The programme is organised under the auspices of a series of Memoranda of 
Understanding (MOU) with over 70 groups who have access to observing resources that can
participate in the follow-up. The MOU maintains confidentiality until the discoveries
are announced.  In the first aLIGO/Virgo observing run (O1)
a number of candidate optical counterparts were identified by various wide field
optical imaging facilities.   In this paper we mainly discuss the work carried out by the 2.0 metre robotic 
Liverpool Telescope
(LT) to spectroscopically follow up a number of those candidates.
We also draw some broad conclusions about a potential spectroscopic ``classification gap''.  
More details of the work presented here can be found in \cite{copper-gw}.

\section{Expected Optical Signatures}

It is a reasonable expectation that binary mergers involving one or more neutron stars (NS) should show
a transient EM signature due to energetic outflows.  An EM signature is less likely for
the merger of two black holes (BH).  For NS+NS and NS+BH mergers we can anticipate a number
of scenarios based on the assumption that the energetic outflow will have a jet configuration:
\begin{itemize}
\item If the observer is within the jet opening angle, a ``prompt'' spectral
signature similar to a short GRB (e.g. \cite[Berger 2014]{berger}) might be expected.

\item If the observer is outside the jet opening angle, then (infra-)red kilo-nova
emission from radioactive decay of heavy elements synthesised in the merger ejecta is 
predicted (e.g. \cite[Li \& Paczy{\'n}ski 1998]{li}).  
Such emission is likely to be delayed with respect to the GW signal.

\item If the jet Lorentz factor is low, we may find a "failed GRB" orphan afterglow
which lacks high energy emission but still has an X-ray/optical/radio signature (\cite[Lamb \& Kobayashi 2016]{lamb}).
\end{itemize}

\section{Methodology}

The aLIGO/Virgo consortium have put in place a number of mechanisms for communication
of information about GW events and the subsequent follow-up activity between the MOU
partners.  These include GCN Notices (machine readable information packets that can
be rapidly distributed), GCN Circulars (human readable information that is distributed
by email), and a machine and human read/write database system (GraceDB) that
distributes information on burst times, localisations (sky maps) and false alarm rates.
All MOU partners are able to use the GCN and GraceDB systems to provide 
information regarding their follow-up activities (e.g. search footprints and candidate
counterparts) to other MOU partners.

The natural follow-up sequence for the search is to use wide
field facilities to identify potential EM counterparts in the area covered by the
GW localisation sky map
followed by more
detailed observations with narrower field instruments.
Examples of some of the optical wide field instruments used during O1 include
iPTF, PanSTARRS, SkyMapper, VISTA, MASTER, TOROS, TAROR, VST, DECam and Pi of the Sky.
At higher energies facilities included Fermi, INTEGRAL, SWIFT.  In the radio regime
MWA, ASKAP and LOFAR were used.   Follow-up of counterparts was carried out by facilities 
including Keck, PESSTO, UH2.2 and LT (optical spectroscopy) and VLA 
(radio).

\section{Results}

\begin{figure}[t]
\begin{center}
 \includegraphics[width=4.4in,angle=270]{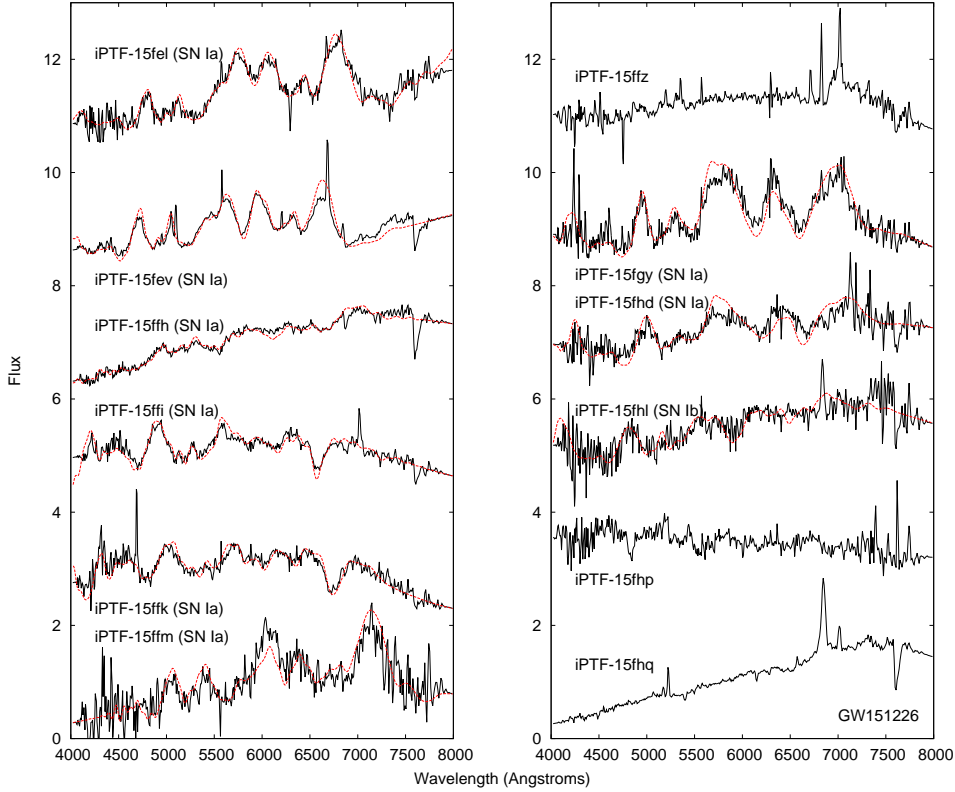} 
 \caption{
 LT spectra obtained during the follow-up of GW151226. 
 We omit candidates for which our observations showed no evidence of a transient. 
 For objects where a supernova identification is obtained
 we overplot the best template fit.}
   \label{fig:gw151226}
\end{center}
\end{figure}

\begin{table}
\caption{Classifications of transients candidates followed by the LT in response to event GW151226. 
Where we obtain a secure supernova classification we provide a redshift, the time $t$ since peak, 
and the percentage of matching templates in the SNID database which are consistent with the spectrum.}
\label{tab:1}
\begin{center}
\begin{tabular}{ll}
Candidate ID & Comments \\
\hline
iPTF-15fed   & No transient detected to limiting magnitude of $R$$\sim$$19.1$\\
iPTF-15fel   & Supernova Type Ia, $z=0.038$, $t=+40$ d, $97.7$ per cent template fit\\
iPTF-15fev   & Supernova Type Ia, $z=0.023$, $t=+50$ d, $94.7$ per cent template fit\\
iPTF-15ffh   & Possible supernova Type Ia, $z=0.061$, $t=+15$d\\
iPTF-15ffi   & Supernova Type Ia, $z=0.085$, $t=+3$ d, $89.1$ per cent template fit\\
iPTF-15ffk   & Supernova Type Ia, $z=0.102$, $t=+5$ d\\
iPTF-15ffm   & Supernova Type Ia, $z=0.094$, $t=+36$ d\\
iPTF-15ffz   & Emission lines consistent with AGN at $z$$\sim$$0.07$\\
iPTF-15fgy   & Supernova Type Ia, $z=0.076$, $t=+20$ d, $84.7$ per cent template fit\\
iPTF-15fhd   & Possible supernova Type Ia, $z=0.091$, $t=+11$ d\\
iPTF-15fhl   & Possible supernova Type Ib, $z=0.043$, $t=+18$ d\\
iPTF-15fhp   & Possible supernova Type Ic, $z=0.129$, $t=+1$ d\\
iPTF-15fhq   & Narrow emission lines, consistent with AGN at $z=0.043$\\
iPTF-15fib   & Slow moving asteroid \\
LSQ15bvw     & No transient detected to limiting magnitude R$\sim$19.5\\
MASTER OTJ020906    &No transient detected to limiting magnitude R$\sim$20\\ 
UGC 1410 transient  &No transient detected. ID'd as minor planet 2 606 Odessa\\
\hline
\end{tabular}
\end{center}
\end{table}

During O1 three GW triggers were distributed to the MOU partners:
\begin{itemize}
\item GW150914 (\cite[Abbott et al. 2016b]{abbott-b}).  This comprised a $36^{+5}_{-4} + 29^{+4}_{-4}$M$_\odot$ black hole 
merger at redshift $z\sim0.09$.  It was detected on 2015 Sept 14 (just before the official start of O1) 
and an alert issued to the MOU partners on 2015 Sept 16.
\item G194575 (\cite[LIGO Scientific Collaboration and Virgo 2015]{ligo}).  This alert was issued on 2015 Oct 22 but retracted on 2015 Nov 11 when
the false alarm rate was recomputed and the event was determined not likely to be a real event.
\item GW151226 (\cite[Abbott et al. 2016c]{abbot-c}).  This comprised a $14.2^{+8.3}_{-3.7} + 7.5^{+2.3}_{-2.3} $M$_\odot$
black hole merger also at redshift $z\sim0.09$.  It was detected on 2015 Dec 26 and an alert
issued on 2015 Dec 27.
\end{itemize}

For all three of these triggers extensive follow-up was carried out.    A summary
of the GW150914 follow-up campaign is presented in \cite{abbott-d}.  Figure \ref{fig:gw151226}
shows a selection of the spectra obtained by the LT
using the low resolution ($R\sim350$) SPRAT spectrograph (\cite{sprat}) in the follow-up of GW151226. A summary
of all of the LT classifications for that event is presented in Table \ref{tab:1}.  Examining the
table shows the majority of candidate counterparts detected are supernovae in the redshift range
$z\sim0.02 - 0.13$.  Some Solar System sources and AGN are also identified, as well as a number
of objects where the transient source had faded back into the galaxy background before
a spectrum could be obtained.  None of the candidates could be associated with GW151226.

\section{Conclusions}

Given that the O1 GW sources were both BH+BH systems (where we do not expect 
an EM signature) the overall results of the follow-up and classification
programme are encouraging.  We have shown that a 2-metre class telescope with a 
high-throughput,
low resolution spectrograph can eliminate many candidate counterparts
at redshifts similar to the GW sources.  In future aLIGO/Virgo observing runs
it is proposed that a distance estimate (``3d sky localisation'' 
- \cite[Singer et al. 2016]{singer}) 
and {\sc ``EM-BRIGHT''} flag (indicating
the possible presence of a neutron star) will be distributed with the alerts.
Localisations will also improve as more GW stations come on-line.  All of this will
help with targeted follow-up of candidates associated with 
potential host galaxies in the correct redshift range.

For GW151226 a total of 77 candidate optical counterparts were announced via GCN.
We note that (over all of the facilities involved):
\begin{itemize}
\item 37 of these received a firm spectral classification,
\item a further 18 had a more tentative classification based on photometric light curves,
\item there were 3 cases where the transient had faded into the host galaxy before 
spectroscopy could be attempted.
\end{itemize}

It follows that 19 candidates were not followed up.  It is therefore
clear that there is a significant danger of a ``classification gap'' opening up,
where potential counterparts will be discovered at a significantly faster rate than
can be spectroscopically followed up.  While localisation error boxes are anticipated
to reduce in size (from the current hundreds of square degrees to tens
of square degrees) as more GW stations come on line, it is also anticipated that event
numbers will increase by an order of magnitude over the next few years as detector 
sensitivities
improve.  The gap
is therefore likely to remain a problem.  In addition we note that new sources of 
transients such as LSST (predicted
to discover $\sim10^6$ transients/night -- \cite[LSST Science collaboration et al. 2009]{lsst}) will also be in competition for 
spectroscopic follow-up time.

To reduce the classification gap we propose the community must (a) start moving
to more automated and efficient methods of triggering spectroscopic follow-up of candidates 
using technologies such
as RTML (\cite[Hessman 2006]{rtml}) and VOEvent (\cite[Williams \& Seaman 2006]{vovenet}) and (b) consider the construction of optimized
spectroscopic follow-up facilities with large apertures and fast slew speeds 
(e.g. Liverpool Telescope 2 - \cite[Copperwheat et al. 2015]{lt2}).

\end{document}